\documentclass[conference]{IEEEtran}
\IEEEoverridecommandlockouts
\usepackage{balance}
\usepackage{cite}
\usepackage{amsmath,amssymb,amsfonts}
\usepackage{algorithmic}
\usepackage{graphicx}
\usepackage{textcomp}
\usepackage{xcolor}
\usepackage{comment}
\usepackage[export]{adjustbox}
\usepackage{multirow}

\newcommand{\ramin}[1]{{\textsf{\textcolor{red}{[From Ramin: #1]}}}}

\def\BibTeX{{\rm B\kern-.05em{\sc i\kern-.025em b}\kern-.08em
    T\kern-.1667em\lower.7ex\hbox{E}\kern-.125emX}}
\begin{document}

\title{Character-based Outfit Generation with Vision-augmented Style Extraction via LLMs
}

\author{\IEEEauthorblockN{Najmeh Forouzandehmehr$^*$, Yijie Cao$^*$, Nikhil Thakurdesai$^*$, Ramin Giahi$^*$, Luyi Ma$^*$, \\ Nima Farrokhsiar, Jianpeng Xu, Evren Korpeoglu, Kannan Achan}

\IEEEauthorblockA{\textit{Personalization Team} \\
\textit{Walmart Global Tech}\\
Sunnyvale, CA, USA \\
\{najmeh.forouzandehmehr, yijie.cao, nikhil.thakurdesai, ramin.giahi, luyi.ma,\\ nima.farrokhsiar, jianpeng.xu, ekorpeoglu, kannan.achan\}@walmart.com}
}


\maketitle

\def\thefootnote{*}\footnotetext{Equal Contribution}\def\thefootnote{\arabic{footnote}}

\begin{abstract}

The outfit generation problem involves recommending a complete outfit to a user based on their interests. 
Existing approaches focus on recommending items based on anchor items or specific query styles but do not consider customer interests in famous characters from movie, social media, etc. 
In this paper, we define a new Character-based Outfit Generation (COG) problem, designed to accurately interpret character information and generate complete outfit sets according to customer specifications such as age and gender. 
To tackle this problem, we propose a novel framework LVA-COG that leverages Large Language Models (LLMs) to extract insights from customer interests (e.g., character information) and employ prompt engineering techniques for accurate understanding of customer preferences. Additionally, we incorporate text-to-image models to enhance the visual understanding and generation (factual or counterfactual) of cohesive outfits. 
Our framework integrates LLMs with text-to-image models and improves the customer's approach to fashion by generating personalized recommendations. 
With experiments and case studies, we demonstrate the effectiveness of our solution from multiple dimensions.
\end{abstract}


\section{Introduction}

\begin{figure}
\includegraphics[width=0.97\linewidth]{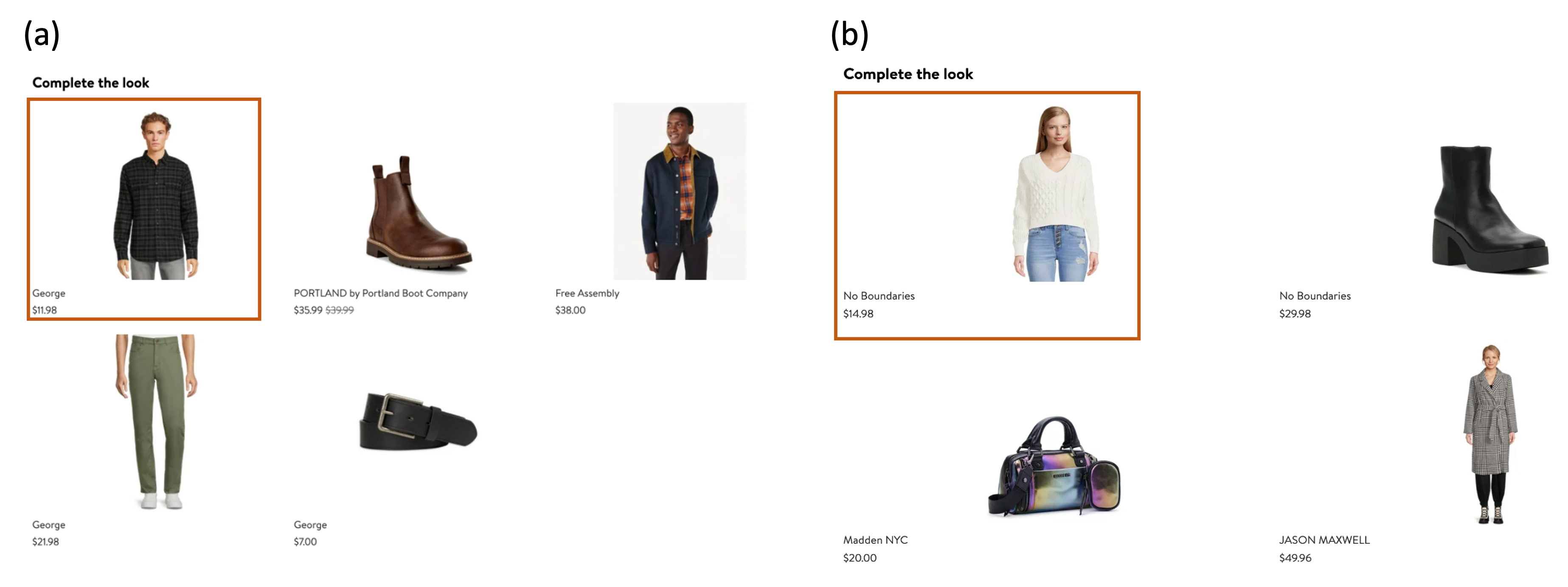}
\caption{Two examples of the outfit generation by the Complete-the-Look (CTL) module in \textit{Walmart.com} which generates the outfit based on a specific item customers have already possessed or have expressed interest in. The anchor items of interest are highlighted in the red box. The rest of items in (a) and (b) are the compatible items to complete the outfit, respectively.} 
\label{fig:walmart_ctl}
\end{figure}

The surging demand for fashion recommendations has presented a multitude of challenges to fashion e-commerce platforms. 
Among these challenges, one of the most pressing is the outfit generation problem which involves the task of recommending a complete outfit to a user, based on their interests \cite{10.1145/3624733}. 
An example of the outfit generation in \textit{Walmart.com}\footnote{https://medium.com/walmartglobaltech/personalized-complete-the-look-model-ea093aba0b73} is shown in Figure \ref{fig:walmart_ctl}, where a customer is interested in a man \textit{long sleeve shirt} and another customer is looking for a woman \textit{v-neck long sleeve sweater}.
For different customers, a e-commerce platform needs to generate the outfit set to complete the look for both items of interest (i.e., anchor items) based on the styles, color and patterns. 
Successfully addressing this problem requires a deep understanding of fashion style, compatibility, and user interest \cite{li2020bootstrapping}. 

To reduce the outfit curation effort and offer more coherent and convenient shopping experience, various approaches have been proposed to recommend items for outfit generation \cite{deldjoo2022review}. 
Many works focus on directly modeling fashion item compatibility between the anchor item and the candidate items \cite{Becattini2023TransformerOG}\cite{Bettaney2020FashionEcommerce}\cite{nakamura2018outfit}
The fill-in-the-blank (FITB) is another approach recommending an item given a missing one in an outfit. The recommended item should match well with the existing outfit \cite{han2017learning}\cite{lorbert2021scalable}.
However, because these methods rely on the co-occurrence of different fashion elements (e.g., style, topic) and items to learn the general compatibility, they all have a common limitation -- 
struggling to provide an effective outfit set when it comes to understanding the customer's interests in a more detailed manner. For instance, they do not address the use cases when a customer wants to dress like a particular movie star or follow a style which is specific to a certain age and gender combination, which requires both domain knowledge, personalization and proper reasoning to infer the style.

\begin{figure}

\includegraphics[width=0.97\linewidth]{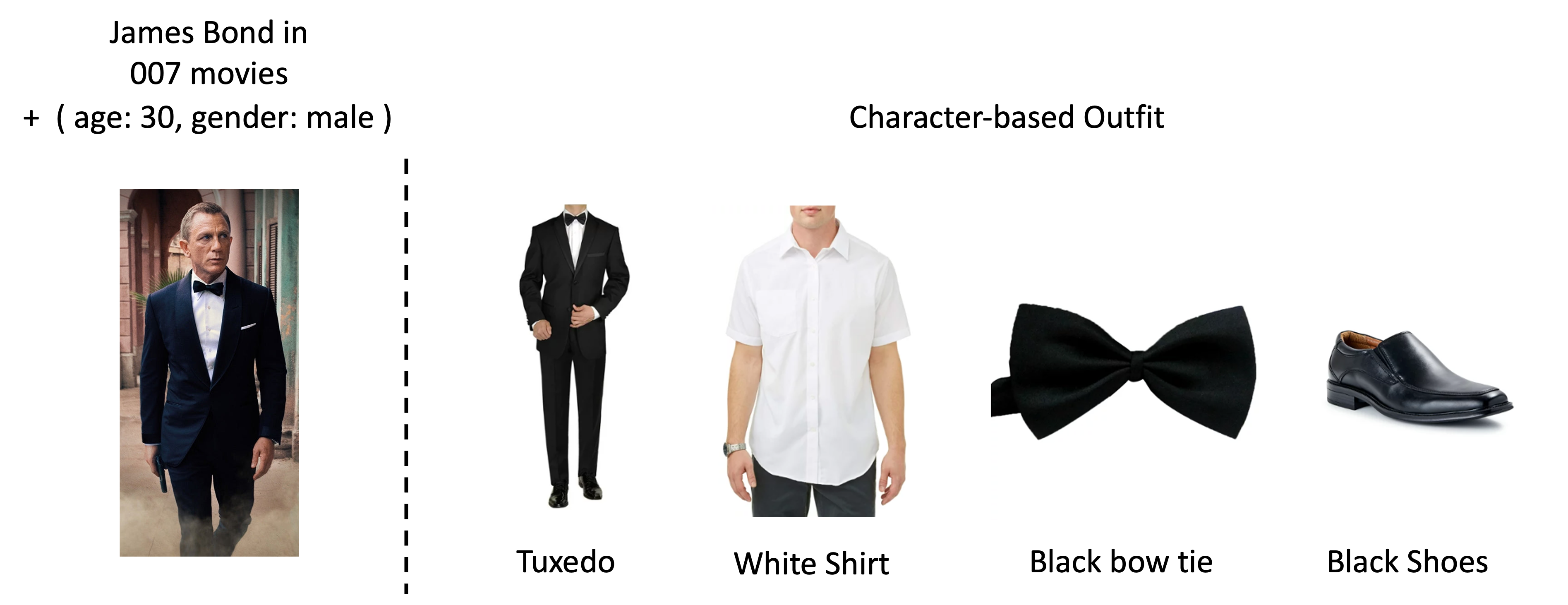}
\caption{Generated outfit (right panel) based on James Bond character (left panel). Note the the image is from Internet and listed here for visualizing the character.}
\label{fig:007}

\end{figure}

In our paper, we define a new task in the outfit generation setting to better address above limitation.
We introduce the \textit{\underline{C}}haracter-based \textit{\underline{O}}utfit \textit{\underline{G}}eneration (COG) task. 
The goal of the COG task is to precisely understand the character information (e.g., style, topic, relevant items and etc.) and generate the complete outfit set given the character's name and customers' specification (e.g., age and gender). 
For example, given a character name (e.g., \textit{James Bond - 007}) with a customer-specified age (30) and gender (male), a set of outfit items illustrated in Figure \ref{fig:007} is expected to solve the COG task to address a traditional English sense of style. 
The COG faces several significant challenges that distinguish it from the traditional outfit generation process. 
Firstly, there is a lack of anchor items and style information that can serve as style and compatibility references, where previous methods heavily rely on these inputs to guide the outfit generation process. 
Secondly, the recommendations need to be tailored to each customer's specific conditions, such as age and gender, which requires the capability of proper reasoning and personalization. 
Lastly, a robust retrieval system needs to be developed to generate recommendation sets with limited information about the character and customer-specific conditions which could be factual or even counterfactual with regard to the classic understanding of the character (e.g., a classic male character with a different age and gender).


Recently, there has been a significant focus on Large-Language Models (LLMs) such as Llama2 \cite{touvron2023llama} and GPT-4 \cite{openai2023gpt4}, and their application in various recommendation problems. 
LLMs are being extensively utilized to analyze customer interests, such as search queries and anchor items, with the aim of enhancing recommendations. They have been successful in incorporating open-domain knowledge into e-commerce scenarios \cite{maragheh2023llm, chen2023knowledge}. However, their potential in outfit generation remains unexplored. 
Additionally, text-to-image models such as CLIP \cite{radford2021learning} and Stable Diffusion (SD) \cite{rombach2022high} have been exploration to map customers' interest to visual space, which address the vision-based relevance in recommender systems \cite{sevegnani2022contrastive}\cite{dong2023musechat}.


To address the challenges of the COG task, we propose a framework called the \textit{\underline{L}}LM-powered \textit{\underline{V}}ision-\textit{\underline{a}}ugmented \textit{\underline{COG}} (LVA-COG) that effectively resolves the aforementioned challenges. 
Our framework leverages the power of Llama2 to gain fashion insights behind customers' interests, addressing concerns related to style and compatibility references. Additionally, we employ prompt-engineering techniques to accurately understand customer preferences and specifications.
Essential item prototypes are inferred by Llama2 to retrieve relevant items. 
However, we acknowledge that outfit generation requires a visual understanding, and relying solely on these two approaches may be insufficient. 
To overcome this limitation, we incorporate SDXL \cite{podell2023sdxl} to visually augment the combination of the character and the customer's (age, gender) pair. 
We keep the image segment of items to retrieve the relevant items and address the visual relevance. 
By integrating LLMs with text-to-iamge models, we enhance the understanding and generation of cohesive outfits.
As different e-commerce platforms have distinct item sets and style of interests, LVA-COG is also designed for robust applications, ready to incorporate the in-house item catalog system. The process of discovering character-based outfits becomes more streamlined, enhancing the overall shopping experience.

Our contributions are summarized as follows:
\begin{itemize}
    \item To the best of our knowledge, this work represents the first definition of the COG task.
    \item Our proposed solution, LVA-COG, not only generates complete outfits without the need for anchor items but also personalizes and tailors the recommendations based on individual customer conditions, including age and gender.
    \item This work pioneers the consideration of LLMs and multi-modality in outfit generation, making it the first to integrate these factors into the process.
\end{itemize}



The paper is structured as follows: Section \ref{literature} presents an overview of related work. In Section \ref{prelim}, we provide a comprehensive description of the problem. Detailed insights into the methodology and architecture of our model are presented in Section \ref{method}. Section \ref{eval} showcases the results of our experiments, including comparisons with baseline models. Finally, Section \ref{conclusion} presents our conclusions and discusses future directions for further research.

\section{Related Work} \label{literature}
\subsection{Outfit Generation} 

Depending on different representations of the customer's interest for outfit generation, previous methods could be classified into three categories. 
Methods in the first category leverage the seed items to represents the customer's interest.
\cite{han2017learning}\cite{Becattini2023TransformerOG}\cite{Bettaney2020FashionEcommerce}\cite{nakamura2018outfit}\cite{lorbert2021scalable}.
They mostly focus on the compatibility among fashion items based on the fashion item co-occurrence in the dataset \cite{mcauley2015image}\cite{veit2015learning}\cite{shih2018compatibility}\cite{vasileva2018learning}
\cite{cucurull2019context}\cite{lin2020fashion}, or the sequence of fashion items in transaction history \cite{han2017learning}\cite{Becattini2023TransformerOG}\cite{sarkar2023outfittransformer}.
The second category involves methods based on query style rather than one or multiple seed items \cite{kang2019complete}\cite{li2019coherent}.
Kang et al. in \cite{kang2019complete} propose a scene-based solution to extract the style of interest, while 
Li et al. in \cite{li2019coherent} focus on the outfit query in natural language with explicit style description.
Methods in the third categories directly model the user interests with personalization \cite{chen2019pog}\cite{ding2023personalized}.
In all three categories, multimodal data of fashion items (e.g., name, category, attributes and images) are commonly used to retrieve the relevant fashion items. 
However, methods in the first two categories heavily rely on the query items or style and mainly learn the compatibility seen in the dataset. 
Methods in the third category fail to incorporate the detailed instruction from customers for outfit generation. 
Our method combines both advantages to clearly infer customers' interest and item compatibilities.

\subsection{Generative Recommender System}
Large Language Models (LLMs) continue to improve the content generation quality and their application in e-commerce \cite{cui2022m6}\cite{geng2022recommendation}\cite{bao2023tallrec}\cite{hou2023large}.
They leverage the capability of LLMs in understanding nature language for better item relevance. 
Compatibility of items is also addressed by \cite{chen2023knowledge} where LLMs are asked to label the relationship (e.g., complementary or substitutable) between two products. 
Besides the text-to-text approach, the text-to-image approach also bridge the gap between text and vision relevance with multimodality input. 
A typical example is the CLIP model \cite{radford2021learning} which encodes the text inputs into latent vectors space via pre-trained language models.
Generative adversarial networks (GANs) are employed in \cite{reed2016generative} to connect text and image modeling, followed by other methods using GANs such as \cite{ruan2021dae
}\cite{zhu2019dm}.
The Stable Diffusion (SD) model in \cite{rombach2022high} and its variants \cite{saharia2022photorealistic}\cite{ruiz2023dreambooth} increases the quality of the text-to-image generation with diffusion probability models (DMs), which inspires 
\cite{sevegnani2022contrastive}\cite{dong2023musechat} for better vision-based recommender systems. 
Our method combines both LLMs and SD in the same framework to streamline the LLM-based preference understanding and vision-based augmentation and relevance. 
\section{Preliminaries}\label{prelim}
\subsection{Problem Definition -- COG}
In this section, we formally define the Character-based Outfit Generation (COG) task. 
Denote $c \in \mathcal{C}$ as a character name, $a \in \mathcal{A}$ as an age, $g \in \mathcal{G}$ as a gender.
Given a triplet of $(c, a, g)$, a fashion item catalog $\mathcal{I} = \{i_1, i_2, i_3, ...\}$, and items' meta-information $\mathcal{M} = \{m_1, m_2, m_3, ...\}$, a COG recommender system is expected to offer a list of compatible fashion items $\mathcal{S} \subseteq \mathcal{I}$ which follow the style behind the character $c$ and match the requirement of the age $a$ and the gender $g$.

Typically, we assume that items' meta-information could contain varied types of data, including attributes such as the item name, color, age, gender, size and etc., and images of items. 
In this paper, the character names were sourced from a curated list of popular characters from movies, TV shows, and literature.
Customer age and gender metadata were also curated based on each character.


\section{Methodology} \label{method}

\begin{figure*}[t]
  \centering
  \includegraphics[width=\linewidth]{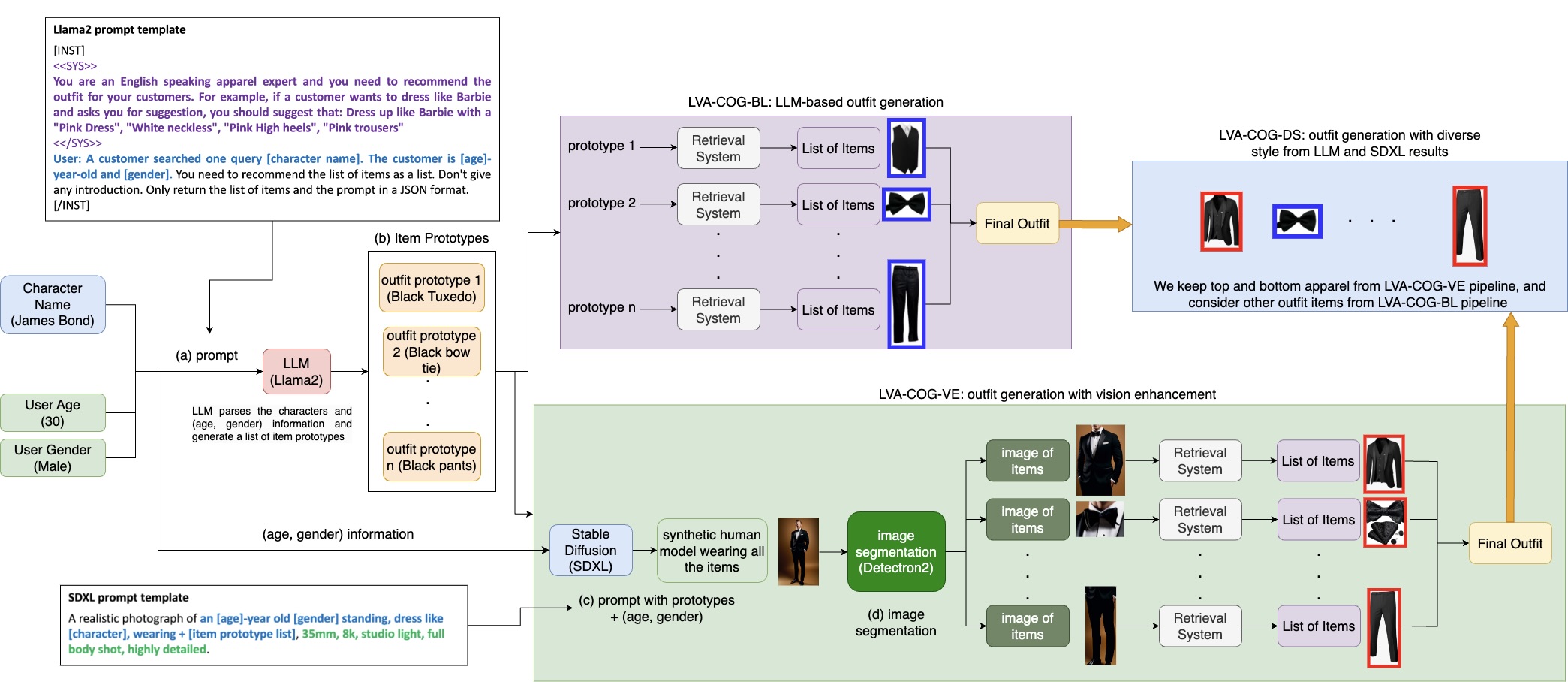}
  \caption{Architecture of our LVA-COG solution. Note that we refer the generic item retrieval system in LVA-COG-BL and LVA-COG-VE  to any implementation based on the item catalog in different e-commerce platforms.}
\label{fig:llm_search_pipeline}
\end{figure*}


In this section, we will introduce our model LVA-COG.
Our method leverages the LLM and the Stable diffusion model, and we propose three variants to generate the outfit set with a (character, age, gender) triplet. 


\subsection{Baseline OG with LLM (LVA-COG-BL)}
In this variant, we only use Llama2 to parse the character of interest and the customer-provide (age, gender) pair. 
We conduct prompt engineering to embed the character, age and gender into a prompt and ask LLMs to generate a set of compatible fashion item prototypes. 
For example, a triplet of (James Bond, 30, male) is provided to an LLM with a prompt (Figure \ref{fig:llm_search_pipeline}-(a)), the LLM will return a black tuxedo, black bow tie, black pants and other prototypes (Figure \ref{fig:llm_search_pipeline}-(b)).
Each prototypes in this set is represented by an item name and a short description.
Given a text-based item-matching engine (e.g., in-house search API), each prototype could be mapped to top relevant items based on the current item catalog in the e-commerce platform. 
The most relevant item is kept for each prototype and a combination of these top relevant items is the final outfit. 
In Figure \ref{fig:llm_search_pipeline}, we highlight the item retrieval step of LVA-COG-BL variant in a purple box. 
We consider the prompt template in Figure \ref{fig:llm_search_pipeline}-(\textit{Llama2 prompt template}) to infer item prototypes from Llama2.




\subsection{Vision-Enhanced OG (LVA-COG-VE)}
Besides the LLM-based COG solution, we also develop the second variant based on SDXL to explore visual relevance in the COG task.
Although multimodality is widely considered to improve outfit generations, our motivation goes beyond that. 
Given the same item prototype, the retrieved items with different (age, gender) pairs are not only different at their item names and attributes, but also the style and designs which are mainly represented visually for a chosen character.
Since our focus in this variant is on the visual relevance and style in item retrieval, we reuse Llama2 to generate a list of item prototypes. 
We leverage SDXL to visualize the item prototypes of a character under different combination of (age, gender), which addresses the subtle requirement of style and design more accurately. 
We highlight the vision-enhancement steps with SDXL in LVA-COG-VE in a green box in Figure \ref{fig:llm_search_pipeline}.

\subsubsection{Step 1 -- Prompt Creation} 
Based on the list of prototypes from the Llama2, we curate a prompt that incorporates (age, gender) accordingly. 
The prompt directs SDXL to generate an artificial human of the specified (age, gender), adorned with the recommended prototypes. Figure \ref{fig:llm_search_pipeline}-(\textit{SDXL prompt template}) shows the prompt template.
A curated and fixed negative prompt is also fed into the model to prevent and unwanted results. 
With these prompts, SDXL generates an image with a human wearing relevant outfit items, which aligns and visualize the combination of (character, age, gender) in visual space.  


\subsubsection{Step 2 -- Image Segmentation}
Based on the output image of SDXL, we leverage a fine-tuned Detectron2 \cite{wu2019detectron2} model on the DeepFashion2 dataset \cite{ge2019deepfashion2} to detect the image segments of fashion items (Figure \ref{fig:llm_search_pipeline}-(d). 
The model processes these images and outputs segmented images. 
Each segmented image ideally showcases a distinct look of an item that aligns with the recommendations tailored to the given character and customer metadata. 
The final result of this segmentation process is a set of images, each highlighting a specific recommended item or look. These segmented images provide a clear visual representation, allowing for easier interpretation and assessment of the recommendations. 


\subsubsection{Step 3 -- Multimodal Item Retrieval} 
Given the segmented images in step 2, and a pair of (age, gender), we can use them an any multimodal item matching engine to find the relevant items.
For example, we could build a multimodal item matching engine with CLIP \cite{radford2021learning} embedding in the item catalog, using both item images and descriptions for training or fine-tuning.
For each segmented image representing an item prototype, we provide the segmented image as the query image along with (age, gender) as query text. 
The retrieval system could be implemented differently based on the application and item catalog.  
The top-\textit{N} items are returned based on the similarity to both the image and the text data. 
Finally, for each character, we select the top-relevant item from each prototype and aggregate them to form a comprehensive outfit.


%

\subsection{Diverse Style OG (LVA-COG-DS)}
The third variant is designed to combine different style distillation from LLM and SD (see Figure \ref{fig:llm_search_pipeline} blue box).
Some item prototypes are represented better by text attributes while others are more precisely interpreted by vision.
The third variant, called Diverse Style OG (LVA-COG-DS), extend the second variant LVA-COG-VE by incorporating the retrieved items from the first variant LVA-COG-BL. 
For top and bottom items such as coat and jeans, we primarily consider the retrieved items from the LVA-COG-VE variant due to better representation of pattern and design. 
For other items which are harder to be detected in an images but essential to complete the outfit such as glasses and bow ties, we leverage the items offered by the LVA-COG-BL variant due to the precise description by LLMs. 
The combined items form the final outfit set. 
\section{Evaluation} \label{eval}
To assess the effectiveness of our proposed methods for character-based look generation, we conducted comprehensive experiments. Given the absence of a publicly available dataset for movie/TV series character outfits, we employed two alternative approaches: one utilizing a Large Language Model (LLM) for automated evaluation and the other involving human evaluators 
to rate the outfits.

\subsection{Experimental Setup}
We generated looks for 29 movie/TV series characters (7 female characters, 22 male characters) using three variations of our purposed model.
Since previous works doesn't fit our problem definition, we compare our three variants (LVA-COG-BL, LVA-COG-VE, LVA-COG-DS) directly against both LLMs and human curators for rating to demonstrate the effectiveness of compatibility reasoning, personalization and tailored style.
We consider the item names and descriptions from the private item catalog for fashion.




\subsection{Evaluators}
\paragraph{LLM Evaluators} To quantify the quality of the generated outfits, we introduced a novel evaluation metric. 
We employed another Language Model, GPT4, to evaluate each outfit. 
The GPT4 is prompted with a standardized evaluation question, and it assigned a score on a scale of 1 to 10, with higher scores indicating better quality. 
Given the meta-information of items 
Below is a prompt template we use for LLM evaluation.\\

\noindent\textbf{GPT4 Evaluator Prompt:}\\
\emph{"Evaluate the following outfit for [\textbf{character name}], based on the clothing items listed below, and give it a score of 1 to 10, with 10 being the best:
[Item 1 name]
[Item 1 description]
[Item 2 name]
[Item 2 description]
[Item 3 name]
[Item 3 description],
Please consider the following factors when scoring the outfit:
Appropriateness for the character and personality of [\textbf{character name}], and Overall aesthetic appeal."}

\paragraph{Human Evaluators}
In addition to the LLM-based evaluation, we further assessed the generated outfits through human evaluation. 
We enlisted the help of three human evaluators who rated the looks based on their appropriateness for the respective characters and overall aesthetic appeal. 
The results of this human evaluation were incorporated into the final evaluation scores from 1 to 10.

\subsection{Results}
The evaluation results of LLM and human evaluators are summarized in Table \ref{tab:1} in terms of mean and standard deviation(stdev) of scores for 29 movie/TV character-based outfits generated using LVA-COG-BL, LVA-COG-VE and LVA-COG-DS, respectively.
LVA-COG-DS which combines the strengths of both LVA-COG-BL and LVA-COG-VE approaches, not only achieves the highest mean score according to our LLM-based evaluation but also received consistently favorable human ratings. 
This alignment between the LLM and human evaluators highlights the effectiveness of incorporating search results from LVA-COG-BL. 
It not only enhances the overall quality of the generated looks but also resonates with human judgment, marking it as the best-case scenario.

To study the bias in outfit generation, we focus on the performance of our methods for different genders. 
We conducted a breakdown of human and LLM evaluations for both male and female character looks. 
Our findings consistently revealed that the models performed better in generating female character looks in most cases, both in terms of LLM-based and human evaluations, as detailed in Table \ref{tab:gender_performance}. This observation indicates the presence of gender bias in the look generation process, highlighting the importance of tailoring character-based looks to the specific character's style and personality, which can vary significantly between genders. An in-depth exploration of this gender bias will be a focus of future work, aiming to address and rectify any potential disparities in character-based outfit generation.

\subsection{Case Study 1 -- LLM Evaluator}
We provide two examples of LLM-generated ratings to illustrate the evaluation process:\\

\noindent\textbf{LLM rating for Chandler Bing's look generated by LVA-COG-BL:}\\
\emph{"
Based on Chandler Bing's character and personality from the TV show "Friends," the outfit consisting of a khaki casual workout zipper t-shirt and black slim fit stretch pants presents a more contemporary and athletic appearance. Chandler's fashion sense typically leans towards casual, comfortable, and slightly sophisticated styles. This outfit offers a casual look, but the athletic elements and slim fit make it slightly different from Chandler's usual wardrobe.
Considering the appropriateness for Chandler Bing, I would give this outfit a score of 6 out of 10. While the outfit is still within the realm of casual wear, it doesn't fully align with Chandler's character and personality."}\\

\noindent\textbf{LLM rating for Chandler Bing's look generated by LVA-COG-DS:}\\
\emph{"
Considering the appropriateness for Chandler Bing, I would give this outfit a score of 9 out of 10. The outfit is a great match for his character, combining casual elements with more refined pieces. The white shirt and black jeans create a classic and balanced look, while the oxfords dress shoes add a touch of formality. The outfit manages to be both comfortable and stylish, which is in line with Chandler's personality. 
"}

We can see that the LLM evaluator gives a higher rating to the outfit generated by LVA-COG-DS, due to better matching on the subtle style and aesthetic elements.


\begin{table}[t!]
\centering
\caption{Mean Evaluation Scores and Standard Deviations for different LVA-COG variants}
\resizebox{0.48\textwidth}{!}{
\begin{tabular}{c|c|c|c|c}
\hline
\textbf{Model Variants} & LLM Evaluator  & LLM Evaluator & Human Evaluators & Human Evaluators \\
& mean scores & scores Stdev & mean scores & scores Stdev \\
\hline
LVA-COG-BL & 6.625 & 2.32 & 4.866 & 2.368 \\
LVA-COG-VE & 6.906 & 1.445 & 6.333 & 1.369 \\
LVA-COG-DS & \textbf{7.975} & \textbf{1.413} & \textbf{7.333} & \textbf{1.143} \\
\hline
\end{tabular}
}
\label{tab:1}
\end{table}



\begin{table}
    \centering
    \caption{Mean Evaluation Scores for Female and Male Characters.}
    \resizebox{0.48\textwidth}{!}{
    \begin{tabular}{l|cc|cc}
    \hline
           \multirow{2}{*}{\textbf{Model Variants}} & \multicolumn{2}{c}{\textbf{Female}} & \multicolumn{2}{c}{\textbf{Male}}  \\
          &     LLM Evaluator &  Human Evaluators &     LLM Evaluator &  Human Evaluators \\
          & (mean score) & (mean score) & (mean score) & (mean score) \\ 
    \hline
       LVA-COG-BL & 6.666 & 5.776 & 6.666 & 5.665 \\
LVA-COG-VE & 7.833 & 6.11 & 6.833 & 6.11 \\
LVA-COG-DS & \textbf{8.163} & \textbf{6.988} & \textbf{7.7} & \textbf{6.553} \\
    \hline
    \end{tabular}    
    }
    \label{tab:gender_performance}
\end{table}

\begin{figure}[t!]
  \centering
  \includegraphics[width=0.9\linewidth]{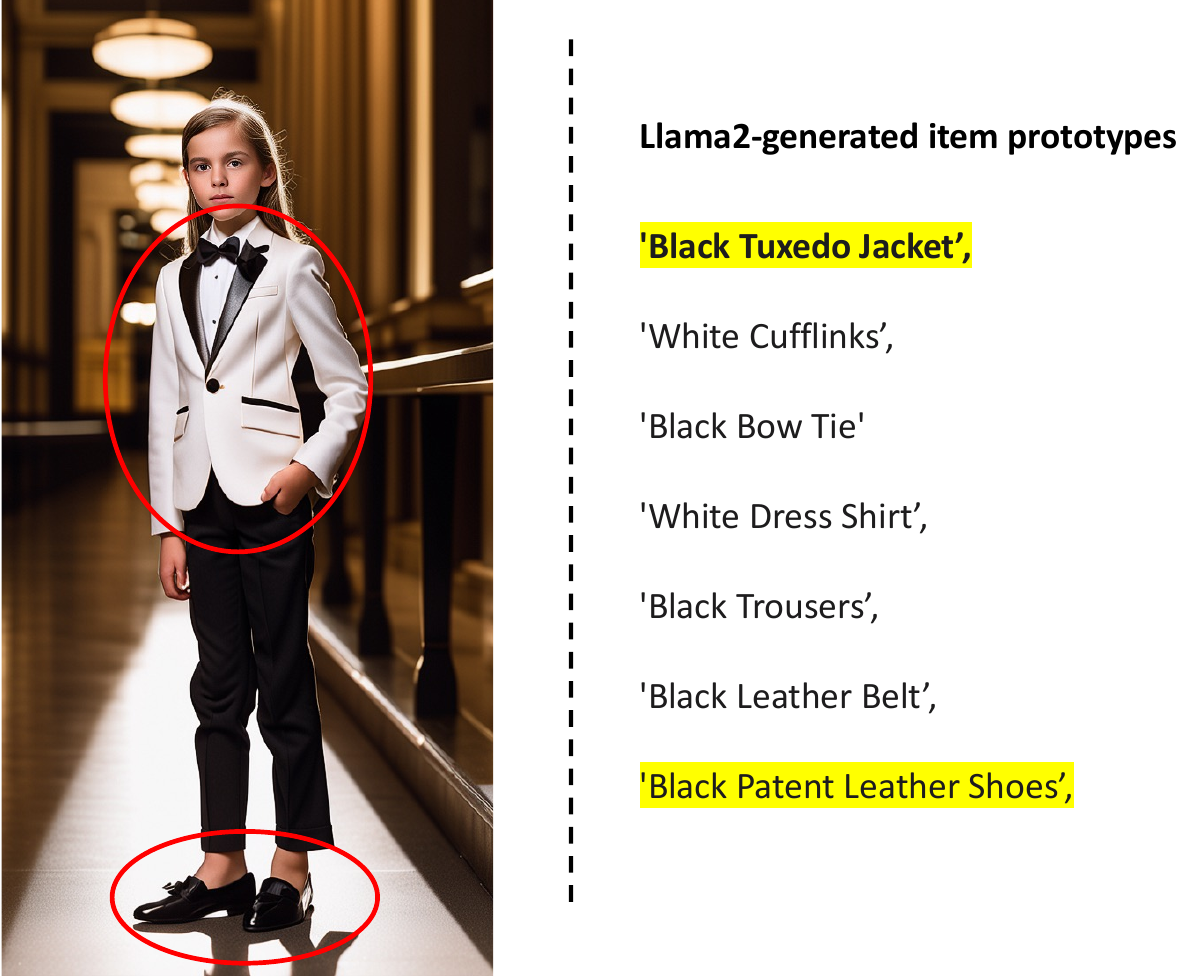}
  \caption{A SDXL-generated image (left panel) and a list of Llama2-generated item prototypes (right panel). The SDXL-generated image creates the novel combination of "James Bond" for a teenage girl and balances the style well compared with the prototypes (highlighted in yellow) which are too general.}
\label{fig:case_study2}
\end{figure}


\subsection{Case Study 2 -- Vision-Enhanced Style Extraction}
To show case the power of our solution in extract the style from the arbitrary combination of (character, age, gender), we ask LVA-COG-VE to generate the synthetic human image wearing all the items for James Bond as a teenage girl (age: 15, gender: female) which is counterfactual compared with the classic understanding of James Bond. 
Such a combination is not existed in the real world. 
Figure \ref{fig:case_study2} compares the SDXL-generated image and the Llama2-generated item prototypes. We can see that the top jacket and the pair of shoes (highlighted in red circles) generated by SDXL address the age and gender well under the English style while the item prototypes by Llama2 is too general to reflect the age and gender.

\subsection{Limitations}

SDXL, while proficient, faces challenges in accurately representing age and gender. For instance, when asked to create a male version of Harley Quinn outfit, the model struggles to adapt, often resulting in images that still portray a female figure. Moreover, the generated images sometimes lack precise details in terms of clothing textures and accessories, which can hinder subsequent processes like image segmentation and vision-based search tasks. To overcome these limitations and improve recommendation quality, advanced prompt engineering techniques can be applied to SDXL and LLM components. Future research could explore incorporating a Multimodal LLM using the character's image as input to enhance the entire pipeline's performance.
\section{Conclusion} \label{conclusion}
In this paper, we propose a new outfit generation task called  Character-based Outfit Generation (COG) and introduce our solution with LLM and Stable Diffusion to infer the style and aesthetic elements.
Evaluations demonstrate the effectiveness of our solution in multiple dimensions.

\bibliographystyle{IEEEtran}
\balance
\bibliography{main} 

\end{document}